\documentclass[10pt]{article}

\usepackage{amsmath}
\usepackage{amssymb}
\usepackage{graphics}
\usepackage{rotating}
\usepackage{cite}
\usepackage{color}
\usepackage{fancybox}

\def\0{\mbox{\tiny $0$}}
\def\1{\mbox{\tiny $1$}}
\def\2{\mbox{\tiny $2$}}
\def\3{\mbox{\tiny $3$}}
\def\4{\mbox{\tiny $4$}}
\def\5{\mbox{\tiny $5$}}
\def\6{\mbox{\tiny $6$}}
\def\7{\mbox{\tiny $7$}}
\def\8{\mbox{\tiny $8$}}
\def\9{\mbox{\tiny $9$}}
\def\I{\mbox{\tiny I}}
\def\II{\mbox{\tiny II}}

\def\mi{\mbox{\tiny $-$}}
\def\pl{\mbox{\tiny $+$}}
\def\ppm{\mbox{\tiny $\pm$}}

\title{\shadowbox{\large \bf DELAY TIME IN QUATERNIONIC QUANTUM MECHANICS}}

\author{
\small  Stefano De Leo\thanks{Department of Applied Mathematics,
State University of Campinas, Brazil [deleo@ime.unicamp.br] } \,\,
and\, Gisele Ducati\thanks{CMCC, Universidade Federal do ABC,
Brasil [ducati@ufabc.edu.br]}}

\date{\small
\fcolorbox{black}{yellow} {\color{red} $\bullet$ {\color{black}{
{\footnotesize  {\sc Journal of Mathematical Physics} {\bf 53}, 022102-8 (2012)}}}
{\color{red}{$\bullet$}} } }

\begin{document}
%
%%%%%%%%%%%%%%%%%%%%%%%%%%%%%%%% PAPER %%%%%%%%%%%%%%%%%%%%%%%%%%%%%%%%%%%%%

\maketitle

\vspace*{-.7cm}

\begin{abstract}
\noindent In looking for quaternionic violations of quantum
mechanics, we discuss the delay time for pure quaternionic
potentials. The study shows in which energy region it is possible
to amplify the difference between quaternionic and complex quantum
mechanics.
\end{abstract}

%%%%%%%%%%%%%%%%%%%%%%%%%%%%%%%%%%%%%%%%%%%%%%%%%%%%%%%%%%%%%%%%%%%%%%%
%%%%%%%%%%%%%%%%%%%%%%%%%%%%%%%%%%%%%%%%%%%%%%%%%%%%%%%%%%%%%%%%%%%%%%%

%%%%%%%%%%%%%%%%%%%%%%%%%%%%%%%%%%%%%%%%%%%%%%%%%%%%%%%%%%%%%%%%%%
%%%%%%%%%%%%%%%%%%%%%%%%%%%%%%%%%%%%%%%%%%%%%%%%%%%%%%%%%%%%%%%%%%%%%%%

%%%%%%%%%%%%%%%%%%%%%%%%%%%%%%%%%%%%%%%%%%%%%%%%%%%%%%%%%%%%%%%%%%%%%%%
%%%%%%%%%%%%%%%%%%%%%%%%%%%%%%%%%%%%%%%%%%%%%%%%%%%%%%%%%%%%%%%%%%%%%%%

%\PACS{ {??.??.?} \and  {??.??.?}{}}
%\PACS{ {42.25.Bs, 42.25.Gy, 42.50.Xa (PACS).}{}}

% Warning: No PACS code given

%02.10.Hh Rings and algebras
%02.10.Ud Linear algebra
%02.10.Yn Matrix theory

%02.30.Hq Ordinary differential equations
%02.30.Jr Partial differential equations
%02.30.Tb Operator theory

%03.65.-w Quantum mechanics
%03.65.Ca Formalism
%03.65.Ta Foundations of quantum mechanics;
%03.65.Xp Tunnelling, traversal time, quantum Zeno dynamics

%12.15.F Quarks and lepton masses and mixing
%14.60.Pq Neutrino mass and mixing

%\offprints{~Stefano De Leo.}

%%%%%%%%%%%%%%%%%%%%%%%%%%%%%%%%%%%%%%%%%%%%%%%%%%%%%%%%%%%%%%%%%%%%%%%
%%%%%%%%%%%%%%%%%%%%%%%%%%%%%%  SECTION   %%%%%%%%%%%%%%%%%%%%%%%%%%%%%
%%%%%%%%%%%%%%%%%%%%%%%%%%%%%%%%%%%%%%%%%%%%%%%%%%%%%%%%%%%%%%%%%%%%%%

\section*{\normalsize I. INTRODUCTION}

Many papers
 have introduced the general framework of
quaternionic quantum
mechanics\cite{DAV89,DAV92,DS00,DD01,DSS02,DDN02,DD03,DD04,DD06,DDLP10}.
The book published in 1995 under the title {\em Quaternionic
quantum mechanics and quantum fields} by Adler\cite{ADL} is,
today, recognized
 to be indispensable to a deeper understanding of the subject.
 It seems, however, that a more
practical side has been somewhat neglected. In other words, the
investigation of quantitative and qualitative differences between
complex and quaternionic quantum mechanics and the consequent
experimental proposals  seem to be in the early stage. It was in
the hope of filling this gap that the authors have recently
reviewed scattering problems in quaternionic quantum
mechanics\cite{DD05,DD06b,DD07}. The more elementary problems,
such as square-well potentials, have been carefully re-considered
with the {\em new} tendency to arrive at more practical results.
The main purpose was to emphasize  quaternionic violations of
complex quantum mechanics which could be tested in laboratory
experiments. This should make the subject more useful to  and
accessible over the worldwide community of scientists interested
in looking for  the existence of quaternionic potentials.

Contrary to what is predicted by classical mechanics, in complex
quantum mechanics a particle with energy $E_{\0}$ is {\em not}
instantaneously reflected by a potential step of height $V_{\0}$
greater than the incoming energy $E_{\0}$\cite{COHEN,MER}. Let us
briefly recall the standard introduction to  delay times.  The
center of the incident wave packet arrives at the potential
discontinuity, $x=0$, at time $t=0$. During a certain interval of
time around $t=0$, the wave packet is localized in the region
where the potential discontinuity is. For sufficiently large times
the incident wave packet  disappears and  the reflected wave
packet  propagates  towards the left at a speed
$\sqrt{2\,E_{\0}/m}$. By using the stationary phase condition
\[
\left[\,\frac{\mbox{d}}{\mbox{d}E}\,\left(\,2\,\theta_c -
\frac{\sqrt{2\,m\,E}}{\hbar}\,\,x_r  -
\frac{E\,t}{\hbar}\right)\,\right]_{\0}=0\,\,,
\]
where $2\,\theta_c$ is  the phase of the reflection coefficient,
 \begin{equation}
 R_{c}=\frac{\sqrt{E}-i\,\sqrt{V_{\0}-E}}{ \sqrt{E}+i\,\sqrt{V_{\0}-E}} =
 \exp[\,2\,i\,\theta_c]\,\,,
\end{equation}
we can calculate the position of the center of the reflected wave
packet,
\[
\sqrt{\frac{m}{2\,E_{\0}}}\,\,x_r=
-\,t+2\,\hbar\,\left[\,\frac{\mbox{d}\theta_c}{\mbox{d}E}\,\right]_{\0}\,\,.
\]
This clearly shows a {\em delay time} in reflection  given by
\begin{equation}
2\,\hbar\,\left[\,\frac{\mbox{d}\theta_c}{\mbox{d}E}\,\right]_{\0}=
\frac{\hbar}{\sqrt{E_{\0}(V_{\0}-E_{\0})}}\,\,.
\end{equation}
This delay time is due to the fact that, for times close to zero,
the probability of presence of the particle in the classically
forbidden region  is {\em not} zero. The study of the delay time
in quaternionic quantum mechanics will be the subject matter of
this paper.

\section*{\normalsize II. ONE-DIMENSIONAL POTENTIALS AND ONE-DIMENSIONAL MOTION}

In order to understand the essential features of quaternionic
quantum mechanics, we focus our attention on quaternionic
one-dimensional problems. They  may be studied deriving the
equation of motion from the three-dimensional quaternionic
Schr\"odinger equation\cite{ADL},
\begin{equation}
\label{q1} \hbar\,\partial_{t} \Phi(\boldsymbol{r},t) = \left[ \,
i\,\frac{\,\,\hbar^{^{2}}}{2m} \, \nabla^{^{2}} - i\,
V_{\1}(\boldsymbol{r}, t) - j \, V_{\2}(\boldsymbol{r}, t) - k \,
V_{\3}(\boldsymbol{r}, t) \, \right] \,
\Phi(\boldsymbol{r},t)\,\,.
\end{equation}
If the potential only depends on one of the three spatial
coordinates, say $x$, and the motion is one-dimensional,
$p_y=p_z=0$,
%only dependence of the potential is on the coordinate $x$, by the following factorization
%\[
%\Phi(\boldsymbol{r},t) = \varphi(x)\, \exp\left[\,i\,\left(p_{\2} \,y + p_{\3}\, z - E\,t %\right)\,/\,\hbar\,\right]\,\,,
%\]
%we can reduce  the partial differential equation (\ref{q1})  to an ordinary differential equation for $\varphi(x)$,
%\begin{equation}
%\label{q2} - E\, \varphi(x)\, i =
%i \, \frac{\hbar^{^{2}}}{2m} \,\varphi''(x)  - i\,\frac{p_{\2}^{\,\2}+p_{\3}^{\,\2}}{2\,m} \,\varphi(x)   - i\, %V_{\1}(x) \,\varphi(x) - j\,V_{\2}(x) \,\varphi(x) - k\, V_{\3}(x)\, \varphi(x)\,\,.
%\end{equation}
%For one-dimensional motion, $p_{\2}=p_{\3}=0$,
the previous equation becomes
\begin{equation}
\label{q3} i \, \frac{\hbar^{^{2}}}{2m} \,\varphi''(x) =
\left[\, i\, V_{\1}(x) + j\,V_q(x)\,e^{-\,i\,\rho(x)}\,\right]
\,\varphi(x) - E\, \varphi(x)\,i\,\,,
\end{equation}
where
\[ V_q(x)=\sqrt{V_{2}^{^{2}}(x)+V_{3}^{^{2}}(x)}\,\,\,\,\,\,\,\,\,\mbox{and}\,\,\,\,\,\,\,
\tan\rho(x)=V_{3}(x)\,/\,V_{2}(x)\,\,.\] Let us now analyze the
stationary states for the case of a quaternionic step potential
defined by
\begin{equation}
\boldsymbol{h}\cdot \boldsymbol{V}(x)
:=\left\{\,\boldsymbol{0}\,\,\,\,\,\mbox{for}\,\, x <
0\,\,\,\,\,\,\,\,\,\mbox{and}\,\,\,\,\,\,\,\,\,\boldsymbol{h}\cdot
\boldsymbol{V}\,\,\,\,\,\mbox{for}\,\,x>0\,\right\}\,\,,
\end{equation}
where $\boldsymbol{h}=(i,j,k)$.  To shorten notation, it is
convenient to introduce the following adimensional quantities
\[
\frac{\sqrt{2m\,V_{\0}}}{\hbar}\,\,x = \xi\,\,\,,\,\,\,\,\,\,\,
\frac{V_q}{V_{\0}} =
\nu_q\,\,\,,\,\,\,\,\,\,\,\mbox{and}\,\,\,\,\,\,\,
 \frac{E}{V_{\0}} = \epsilon\,\,,
\]
where $V_{\0}=\sqrt{V_{\1}^{^{2}}+V_{q}^{^{2}}}$. In terms of this
new adimensional quantities, the differential equation for the
wave function in the free potential region $\xi <0$ becomes
\begin{equation}
\label{s1} i\,\varphi_{\I}''(\xi)= - \epsilon
\,\varphi_{\I}(\xi)\,i\,\,.
\end{equation}
%The general quaternionic solution of Eq.(\ref{s1}) is
%\[ \varphi_{\I}(\xi)= A_{\I}\exp[\,i\,\sqrt{\epsilon}\,\xi\,] + B_{\I}\exp[\,-\,i\,\sqrt{\epsilon}\,\xi\,]+ j\,
%C_{\I}\exp[\,\sqrt{\epsilon}\,\xi\,] + j\,D_{\I}\exp[\,-\,\sqrt{\epsilon}\,\xi\,]\,\,,
% \]
%where $A_{\I}$, $B_{\I}$, $C_{\I}$ and $D_{\I}$ are complex coefficients\cite{DD01}.
For the case of incident particles coming from the left,
% and to guarantee
%bounded solutions when $\xi \to -\,\infty$, we have to set respectively
%$A_{\I}=1$ and $D_{\I}=0$. Renaming $B_{\I}$ by $R$ and  $C_{\I}$ by $\widetilde{R}$,
the free potential plane wave solution is the region $\xi <0$ is
\cite{DD01,DD06b}
\begin{equation}
\label{sol1} \varphi_{\I}(\xi)= \exp[\,i\,\sqrt{\epsilon}\,\xi\,]
+ R\,\exp[\,-\,i\,\sqrt{\epsilon}\,\xi\,]+ j\,
\widetilde{R}\,\exp[\,\sqrt{\epsilon}\,\xi\,]\,\,,
 \end{equation}
 where $R$ and $\widetilde{R}$ are complex coefficients to be determined by the
 matching conditions.
The wave function in the potential region, $\xi >0$, satisfies the
following differential equation
\begin{equation}
\label{s2} i\,\varphi_{\II}''(\xi)= \left( i\, \sqrt{1 -
\nu_q^{\,\2}} + j \, \nu_{q}\,e^{-i\rho}
\right)\,\varphi_{\II}(\xi) - \epsilon
\,\varphi_{\II}(\xi)\,i\,\,,
\end{equation}
whose solution is\cite{DD01,DD06b}
%\[ \varphi_{\II}(\xi) =   ( 1 + j\, \gamma) \,  \left\{
%A_{\II}\exp \left[ \,i\,\alpha_{\mi} \, \xi \, \right] + B_{\II} \exp \left[
%\,-\,i\, \alpha_{\mi} \, \xi \, \right]  \right\} +  ( \beta +
%j ) \, \left\{ C_{\II} \exp \left[ \, - \,\alpha_{\pl} \, \xi \, \right] +
%\exp \left[  \, \alpha_{\pl} \, \xi \, \right]
%\, D_{\II} \right\}\,\,,
%\]
\begin{equation}
\label{sol2} \varphi_{\II}(\xi) =   ( 1 + j\, \gamma) \,\, T\,\exp
\left[ \,i \,\alpha_{-} \, \xi \, \right] +  ( \beta + j ) \,\,
\widetilde{T}\, \exp \left[ \, - \,\alpha_{\pl} \, \xi \, \right]
\,\,.
\end{equation}
where
\begin{equation}
\alpha_{\pm}  =  \sqrt{ \sqrt{\epsilon^{\2} - \nu_{q}^{\,\2}}\,\pm
\sqrt{1 - \nu_q^{\,\2}}} \,\,\,,\,\,\,\,\,
\beta=i\,\frac{\nu_q\,e^{i\rho}}{ \epsilon + \sqrt{\epsilon^{\2} -
\nu_{q}^{^{2}}}}\,\,\,,
\,\,\,\,\,\gamma=-\,i\,\frac{\nu_q\,e^{-i\rho}}{\epsilon +
\sqrt{\epsilon^{\2} - \nu_{q}^{\,\2}}}\,\,\,,
\end{equation}
%and  $A_{\II}$, $B_{\II}$, $C_{\II}$ and $D_{\II}$ are complex coefficients\cite{DD01,DD06}.\\
and $T$ and $\widetilde{T}$ are complex coefficients to be determined by the matching
conditions. \\

\noindent For diffusion, $E>V_{\0}$, we have $\alpha_{\ppm}\in
\,\mathbb{R}_{\pl}$. For the case of total reflection, $E<V_{\0}$,
we find
\begin{eqnarray*}
\nu_q<\epsilon<1 &:& \alpha_{\mi}  =  i\,\sqrt{  \sqrt{1 -
\nu_q^{\,\2}} - \sqrt{\epsilon^{\2} -
\nu_{q}^{\,\2}}} \, \in \, i\,\mathbb{R}_{\pl}\,\,\,\,\,\mbox{and}\,\,\,\,\, \alpha_{\pl}\,\in\,\mathbb{R}_{\pl}\,\,, \\
0<\epsilon<\nu_q & : &  \alpha_{\mi}  = i\, \left(1 -
\epsilon^{\2}\right)^{^{1/4}} \exp\left[\,
-\,\frac{\,i}{2}\,\arctan \sqrt{\frac{\nu_{q}^{\,\2} -
\epsilon^{\2}}{1 - \nu_q^{\,\2}}}\,\right]\,\in
\,\mathbb{C}_{\pl}\,\,\,\,\mbox{and}\,\,\,\,\,\alpha_{\pl}=i\,\alpha_{\mi}^*\,\,.
\end{eqnarray*}
% Observing that $\mbox{Re}[\,i\,\alpha_{\mi}] \in \mathbb{R}_{\mi}$ and  $\mbox{Re}[\,\alpha_{\pl}] \in %\mathbb{R}_{\pl}$, to guarantee bounded solutions when $\xi \to \infty$, we have to impose $B_{\II}=0$ and %$D_{\II}=0$. The presence of evanescent solutions in the potential region  implies total reflection.\\

%\noindent
%Renaming $A_{\II}$ by $T$ and $C_{\II}$ by $\widetilde{T}$, the quaternionic plane wave solution in the potential %region becomes
%\begin{equation}
%\label{sol2}
%\varphi_{\II}(\xi) =   ( 1 + j\, \gamma) \,\,
%T\,\exp \left[ \,i \,\alpha_{-} \, \xi \, \right]
%+  ( \beta +
%j ) \,\, \widetilde{T}\, \exp \left[ \, - \,\alpha_{\pl} \, \xi \, \right] \,\,.
%\end{equation}

\section*{\normalsize III. TOTAL REFLECTION: PHASE CALCULATION}

To obtain the reflection coefficient, we have to use the matching
conditions at $x=0$, i.e.   $\varphi_{\I}(0)=\varphi_{\II}(0)$ and
$\varphi'_{\I}(0)=\varphi'_{\II}(0)$. These conditions lead to the
following quaternionic system,
\begin{eqnarray}
\label{mc}
1+R+j\,\widetilde{R}& =& (1+j\,\gamma)\,\,T + (\beta+j)\,\,\widetilde{T}\,\,,\nonumber \\
\sqrt{\epsilon}\,\left[\,i\,(R-1) - j\,\widetilde{R}\,\right] &=&
-\,(1+j\,\gamma)\,\,T\,i\,\alpha_{\mi} +
(\beta+j)\,\widetilde{T}\,\,\alpha_{\pl}\,\,.
\end{eqnarray}
Multiplying the first equation for $\sqrt{\epsilon}$ and then
summing it with the second equation, we find
\[ \sqrt{\epsilon}\,\,[\,1+R+i\,(R-1)\,] = (1+j\,\gamma)\,\,T\,(\sqrt{\epsilon}-i\,\alpha_{\mi}) + (\beta+j)\,\,\widetilde{T}\,(\sqrt{\epsilon}+\alpha_{\pl})\,\,.\]
The left hand of this equation is complex. Consequently,  the pure
quaternionic part of the right hand equation has to be zero. This
immediately implies the following relation between $T$ and
$\widetilde{T}$,
\[ \widetilde{T}= -\, \gamma\,\,\frac{\sqrt{\epsilon} -i\,
\alpha_{\mi}}{\sqrt{\epsilon}+\alpha_{\pl}}\,\,T\,\,.
\]
By using this relation in Eqs.(\ref{mc}) and taking the ratio of
the complex parts,
%we obtain
%\[
%\frac{1+R}{1-R} = \frac{\sqrt{\epsilon} \,(\sqrt{\epsilon}+\alpha_{\pl}) - \beta \gamma \,\sqrt{\epsilon} %\,(\sqrt{\epsilon}-i\,\alpha_{\mi})}{\alpha_{\mi} (\sqrt{\epsilon}+\alpha_{\pl}) -
%i\, \beta \gamma \,\alpha_{\pl} \,(\sqrt{\epsilon}-i\,\alpha_{\mi}) }\,\,.
%\]
after simple algebraic manipulations, we find the following
reflection coefficient
\begin{equation}
R= \frac{(\sqrt{\epsilon} + \alpha_{\pl})(\sqrt{\epsilon} -
\alpha_{\mi}) -\beta \gamma\,(\sqrt{\epsilon} -i\, \alpha_{\mi})(
\sqrt{\epsilon} - i\, \alpha_{\pl})}{(\sqrt{\epsilon} +
\alpha_{\pl})(\sqrt{\epsilon} +  \alpha_{\mi}) -\beta
\gamma\,(\sqrt{\epsilon} -i\,\alpha_{\mi})( \sqrt{\epsilon} + i\,
\alpha_{\pl})}\,\,.
\end{equation}
We now explicitly calculate the phase for each of the two cases
which characterize total reflection. Let us begin with
$\nu_q<\epsilon<1$. In this case, we have $(\beta \gamma)^*=\beta
\gamma$, $\alpha_{\mi}^*=-\,\alpha_{\mi}$ and
$\alpha_{\pl}^*=\alpha_{\pl}$. This implies
\begin{equation*}
R_{_{>}}(\nu_q):=R(\nu_q<\epsilon<1)=\exp \left[\,2\, i\,
\theta_{_{>}}(\nu_q) \right]\,\,,
\end{equation*}
where
\begin{equation}
\label{thetaup} \theta_{_{>}}(\nu_q)=\arctan\,\left[\, \frac{\beta
\gamma\, \alpha_{\pl} ( \sqrt{\epsilon} -i\, \alpha_{\mi}) +i\,
\alpha_{\mi} (\sqrt{\epsilon} + \alpha_{\pl})}{\sqrt{\epsilon}\, (
\sqrt{\epsilon} + \alpha_{\pl}) - \beta \gamma \,\sqrt{\epsilon}\,
( \sqrt{\epsilon} -i\, \alpha_{\mi})} \right]\,\,.
\end{equation}
The case $0<\epsilon<\nu_q$ is a little bit more complicated. Let
us first observe that $\beta\gamma=\exp[-2\,i\,\omega]$ with $\tan
\omega= \sqrt{\nu_{q}^{\,\2} - \epsilon^{\2}}/\epsilon$ and recall
that in this case $\alpha_{\pl}=i\,\alpha_{\mi}^{*}$.  We then
find
\begin{equation*}
R_{_{<}}(\nu_q):=R(0<\epsilon<\nu_q) = - \,\exp \left[\,
2\,i\,\theta_{_{<}}(\nu_q)\,\right]\,\,,
\end{equation*}
where
\begin{equation}
\label{thetadown} \theta_{_{<}}(\nu_q)   =
\arctan\left(\,\frac{\epsilon\,\sin \omega + \sqrt{\epsilon}\,\,\,
\mbox{Im}\left[\,i\, \alpha_{\mi}^*\, e^{i \omega}
\,\right]}{|\alpha_{\mi}|^{^2} \sin \omega - \sqrt{\epsilon}\,\,\,
\mbox{Re}\left[\,\alpha_{\mi}\,e^{i \omega}\,\right]
}\,\right)\,\,.
\end{equation}
The phase for a {\em pure complex potential} (standard quantum
mechanics) is obtained from Eq.(\ref{thetaup}) by setting
$\nu_q=0$,
\begin{equation}
\label{thetac} \theta_c:=\theta_{_{>}}(0)=-\,\arctan
\sqrt{\frac{1-\epsilon}{\epsilon}}
\end{equation}
For a {\em pure quaternion potential}, by using
Eq.(\ref{thetadown}) for $\nu_q = 1$, we find
%Observing that in this limit, we have
%\[ \alpha_{\mi} = (1 + i)\, (1 - \epsilon^{\2})^{^{1/4}}\,/\,\sqrt{2} = %(1+i)\,\alpha\,\,\,\,\,\mbox{and}\,\,\,\,\,\tan \omega \to 2\,\alpha^{\2}\,/\,\epsilon\,\,,\]
%we find
\begin{equation}
\label{thetaq} \theta_q:=\theta_{_{<}}(1) = \arctan
\left(\,\frac{2\, \alpha\, \epsilon +
2\,\alpha^{\2}\sqrt{\epsilon}  + \epsilon^{^{3/2}}}{ 4\,
\alpha^{^3} + 2\,\alpha^{\2}\sqrt{\epsilon}  -
\epsilon^{^{3/2}}}\,\right)\,\,.
\end{equation}

\section*{\normalsize IV. TOTAL REFLECTION: DELAY TIME CALCULATION}

The quaternionic wave function which determines the particle
dynamics in region I, $\xi<0$, is given by
\begin{equation}
\varphi_{\I}(\xi,\tau)= \left(\,
\underbrace{\exp[\,i\,\sqrt{\epsilon}\,\xi\,]}_{\displaystyle{\varphi_{\I,inc}}}
+ \, \underbrace{R\,\exp[\,-\,i\,\sqrt{\epsilon}\,\xi\,]+ j\,
\widetilde{R}\,\exp[\,\sqrt{\epsilon}\,\xi\,]}_{\displaystyle{\varphi_{\I,ref}}}\,\right)\,
\exp[\,-\,i\,\epsilon\,\tau\,] \,\,,
\end{equation}
where $\tau=V_{\0}\,t/\hbar$. As observed in the introduction, for
sufficiently large times the incident wave packet
($\varphi_{\I,inc}$)  disappears and  the reflected wave packet
($\varphi_{\I,ref}$)  propagates  towards the left. Observing that
the pure quaternionic part decreases exponentially, the stationary
phase method can be directly applied to the complex part of the
reflected quaternionic wave function,
\[
\left[\,\frac{\mbox{d}}{\mbox{d}\epsilon}\,\left(\,2\,\theta -
\sqrt{\epsilon}\,\,\xi_r  -
\epsilon\,\tau\,\right)\right]_{\0}=0\,\,.
\]
The {\em delay time} is then given by
\begin{equation}
\label{dt} \tau_{\0}=
2\,\left[\,\frac{\mbox{d}\theta}{\mbox{d}\epsilon}\,\right]_{\0}\,\,.
\end{equation}
From this equation, by using the phase given in Eq.(\ref{thetac}),
we obtain the the {\em standard} delay time for complex
potentials, i.e.
\begin{equation}
\label{dtc}
\tau_{\0,c}[\epsilon_{\0,c}]=2\,\left[\,\frac{\mbox{d}\theta_c}{\mbox{d}\epsilon}\,\right]_{\0}=
\frac{1}{\sqrt{\epsilon_{\0,c}\,(1-\epsilon_{\0,c})}}\,\,,
\end{equation}
where $\epsilon_{\0,c}=E_{\0}/V_{\1}$. The dependence of
$\tau_{\0,c}=V_{\1}t_{\0}/\hbar$  upon  the ratio between the
incoming energy, $E_{\0}$, and the complex potential, $V_{\1}$, is
plotted in Fig.\,1 (continuous line). The minimum in the plane
$\epsilon_{\0,c}$-$\tau_{\0,c}$ is found at
\begin{equation}
\left\{\,\frac{\widetilde{E}_{\0,c}}{V_{\1}}\,,\,\frac{V_{\1}\,\widetilde{t}_{\0,c}}{\hbar}\,\right\}
= \left\{\,0.5\,,\,2\,\right\}\,\,.
\end{equation}
 Consequently,
\begin{equation}
\left\{\,\frac{\widetilde{E}_{\0}\,\widetilde{t}_{\0}}{\hbar}\,\right\}_c
=\,1\,\,.
\end{equation}
For a {\em pure quaternionic potential}, we have to use the phase
given in Eq.(\ref{thetaq}). The expression for the quaternionic
delay time is a little bit more complicated,
\begin{equation}
\label{dtq}
\tau_{\0,q}[\epsilon_{\0,q}]=2\,\left[\,\frac{\mbox{d}\theta_q}{\mbox{d}\epsilon}\,\right]_{\0}=
\frac{2+\displaystyle{\frac{1}{\sqrt{2\,\epsilon_{\0,q} }\,
\left(1-\epsilon_{\0,q}^{\2}\right)^{^{3/4}}}}+\frac{2\,
\epsilon_{\0,q}
}{\left(1-\epsilon_{\0,q}^{\2}\right)^{^{1/2}}}+\frac{2\,
\sqrt{2\,\epsilon_{\0,q}}}{\left(1-\epsilon_{\0,q}^{\2}\right)^{^{1/4}}}}{\epsilon_{\0,q}
+\sqrt{2\,\epsilon_{\0,q}}
\,\left(1-\epsilon_{\0,q}^{\2}\right)^{^{1/4}}+\left(1-\epsilon_{\0,q}^{\2}\right)^{^{1/2}}}\,\,,
\end{equation}
where $\epsilon_{\0,q}=E_{\0}/V_q$.  The plot of
$\tau_{\0,q}=V_{q}t_{\0}/\hbar$ as a function of the ratio
 between the  incoming energy, $E_{\0}$, and the modulus of the pure quaternionic potential, $V_{q}$,
 is shown in Fig.\,1 (dotted line). Due to the fact that the minimum of the  quaternionic delay time,
\begin{equation}
\left\{\,\frac{\widetilde{E}_{\0,q}}{V_{q}}\,,\,\frac{V_{q}\,
\widetilde{t}_{\0,q}}{\hbar}\,\right\} =
\left\{\,0.365\,,\,2.763\,\right\}\,\,,
\end{equation}
is different from the complex case,  it seems simple to recognize
a pure quaternionic potential by calculating the quantity
$\widetilde{E}_{\0}\,\widetilde{t}_{\0} /\hbar$.  Surprisingly,
 \begin{equation}
\left\{\,\frac{\widetilde{E}_{\0}\,\widetilde{t}_{\0}}{\hbar}\,\right\}_q
\approx \,1.0085\,\,,
\end{equation}
which is very close to the result found for the complex case.
Thus, an experiment involving this measurement does  not represent
the best choice to see quaternionic potentials.

\section*{\normalsize V. CONCLUSIONS}
As observed in the previous section, a calculation of
$\widetilde{E}_{\0}\,\widetilde{t}_{\0} /\hbar$
 done by two observers which are respectively working with a complex and a pure quaternionic  potential step practically gives the same result. In looking for the energy region which could amplify the difference between complex and quaternionic quantum mechanics, it is interesting to introduce new energy and time variables defined in terms of  the incoming energy, $\widetilde{E}_{\0}$, for which
 we have a minimal delay time. In terms of these new variables,
 \[ \widetilde{\epsilon}_{\0}=E_{\0}/\widetilde{E}_{\0}\,\,\,\,\,\mbox{and}\,\,\,\,\,
   \widetilde{\tau}_{\0}=\widetilde{E}_{\0}\,t_{\0}\,/\,\hbar\,\,,\]
 we find
 \begin{eqnarray}
 \epsilon_{\0,c} & = & \frac{\widetilde{E}_{\0}}{V_{\1}}\,\cdot\,\frac{E_{\0}}{\widetilde{E}_{\0}}  = 0.5\,\,\widetilde{\epsilon}_{\0}\,\,,\nonumber \\
 \widetilde{\tau}_{\0,c} & = & \frac{\widetilde{E}_{\0}}{V_{\1}}\,\cdot\,\tau_{\0,c}[ 0.5\,\,\widetilde{\epsilon}_{\0}]=  0.5\,\,\tau_{\0,c}[0.5\,\,\widetilde{\epsilon}_{\0}]\,\,,
 \end{eqnarray}
 and
 \begin{eqnarray}
 \epsilon_{\0,q} & = & \frac{\widetilde{E}_{\0}}{V_{q}}\,\cdot\,\frac{E_{\0}}{\widetilde{E}_{\0}}  = 0.365\,\,\widetilde{\epsilon}_{\0}\,\,,\nonumber \\
 \widetilde{\tau}_{\0,q} & = & \frac{\widetilde{E}_{\0}}{V_{q}}\,\cdot\,\tau_{\0,q}[ 0.365\,\,\widetilde{\epsilon}_{\0}]=  0.365\,\,\tau_{\0,q}[ 0.365\,\,\widetilde{\epsilon}_{\0}]\,\,.
 \end{eqnarray}
Fig.\,2 clearly shows that the best choice to amplify the
difference between complex and pure quaternionic potentials is
achieved for incoming energies closed to $2\,\widetilde{E}_{\0}$
(this energy value represents the limit case between  tunneling
and diffusion for complex potentials). It is also interesting to
observe that a complex potential {\em cannot} completely mimic a
pure quaternionic potential. Indeed, as shown in Fig.\,2 due to
the different shapes of the delay time curves for complex and pure
quaternionic potentials, a complex potential  can  perfectly mimic
a pure quaternionic potential at most in two cases (see
intersections between the two curves shown in Fig.\,2). In
Table\,1, we explicitly shown the delay times for reflection as
function of the incoming energy (first column) for a pure
quaternionic potential of height $20\,\mbox{KeV}$ (second column).
In such a table, we also find the reflection delay times for
complex potentials of height $18.5\,\mbox{KeV}$ (third column),
$16.5\,\mbox{KeV}$ (fourth column) and $14.5\,\mbox{KeV}$ (fifth
column). These complex potentials  only mimic the quaternionic
potential respectively for incoming energy of $2$, $14$ and $8$
KeV. This clearly shows that the reflection delay times obtained
in presence of a quaternionic potential step cannot be reproduced
by a complex potential step. The energy dependence of the delay
times in reflection by a step potential can be then used to
determine whether the step as a quaternionic nature.

The study presented in this paper represents a preliminary
analysis of quaternionic delay times. Further investigations imply
a generalization from plane waves to wave packets and/or from
one-dimensional to three-dimensional problems.

\section*{\small \rm ACKNOWLEDGEMENTS}

The authors thank an anonymous referee for his interesting
comments and useful suggestions. One of the authors (SdL) also
thanks the CNPq ({\em Bolsa de Produtividade em Pesquisa 2010/13})
and the FAPESP ({\em Grant No.\,10/02213-3}) for financial
support.

\newpage

\begin{table}
\begin{center}
\begin{tabular}{|r|r||r|r|r|}\hline
 & $V_q=20$\,KeV &  $V_{\1}=18.5$\,Kev   &  $V_{\1}=16.5$\,Kev & $V_{\1}=14.6$\,Kev \\ \hline
 $E_{\0}$/KeV & KeV $t_{\0}/\,\hbar$ &  KeV $t_{\0}/\,\hbar$ &
  KeV $t_{\0}/\,\hbar$ &   KeV $t_{\0}/\,\hbar$  \\ \hline \hline
  2 & 0.174 & $\star$ 0.174 & 0.186 & 0.199\\ \hline
  4 & 0.147 & 0.131 & 0.141 & 0.154 \\ \hline
  6 & 0.139  & 0.115 & 0.126 & 0.140 \\ \hline
  8 & 0.138 &  0.109 & 0.121 &  $\star$ 0.138 \\ \hline
  10 & 0.142 &  0.108 &  0.124 & 0.147 \\ \hline
  12 & 0.152 & 0.113  & 0.136 & 0.179 \\ \hline
  14 & 0.169 & 0.126  & $\star$ 0.169 & 0.345 \\ \hline
  \end{tabular}
  \end{center}
  \caption{Reflection delay times for a pure quaternionic potential of height
  $V_q=20$\,Kev (second column). The complex potentials (third,
  fourth and fifth column) only mimic the quaternionic potential
  for particular incoming energies (starred entries).
  Consequently, a complex potential cannot reproduce the quaternionic
  results.}
    \end{table}

\newpage

\begin{figure}[hbp]
\hspace*{-1.35cm}
\includegraphics[width=16.5cm, height=22cm, angle=0]{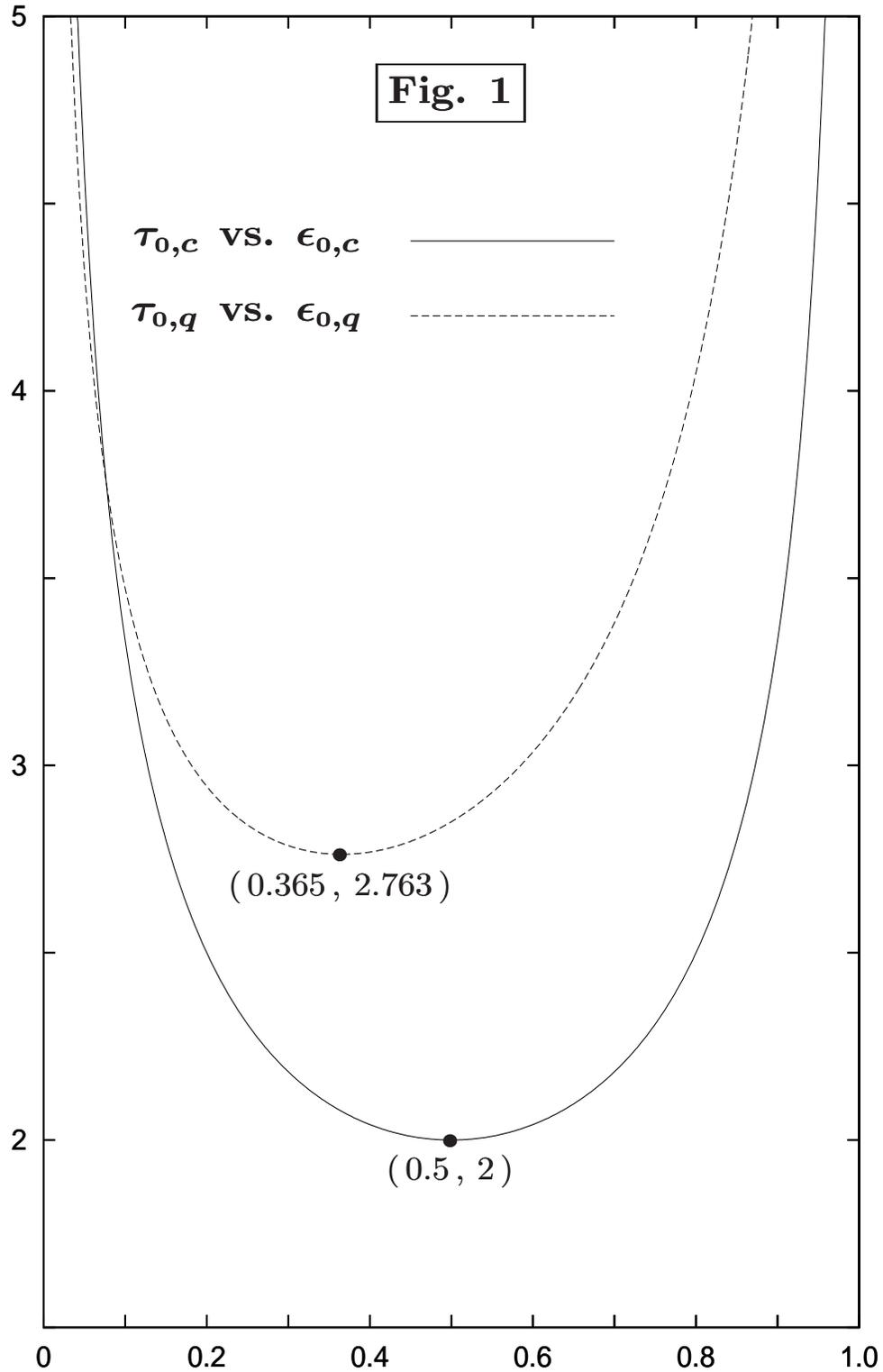}
\vspace*{-1.2cm}
 \caption{Delay times for complex  ($\tau_{\0,c}$) and pure quaternionic  ($\tau_{\0,q}$)  potentials in terms of the adimensional incoming energies $E_{\0}/V_{\1}$ ($\epsilon_{\0,c}$) and $E_{\0}/V_q$ ($\epsilon_{\0,q}$). }
\end{figure}

\newpage

\begin{figure}[hbp]
\hspace*{-1.35cm}
\includegraphics[width=16.5cm, height=22cm, angle=0]{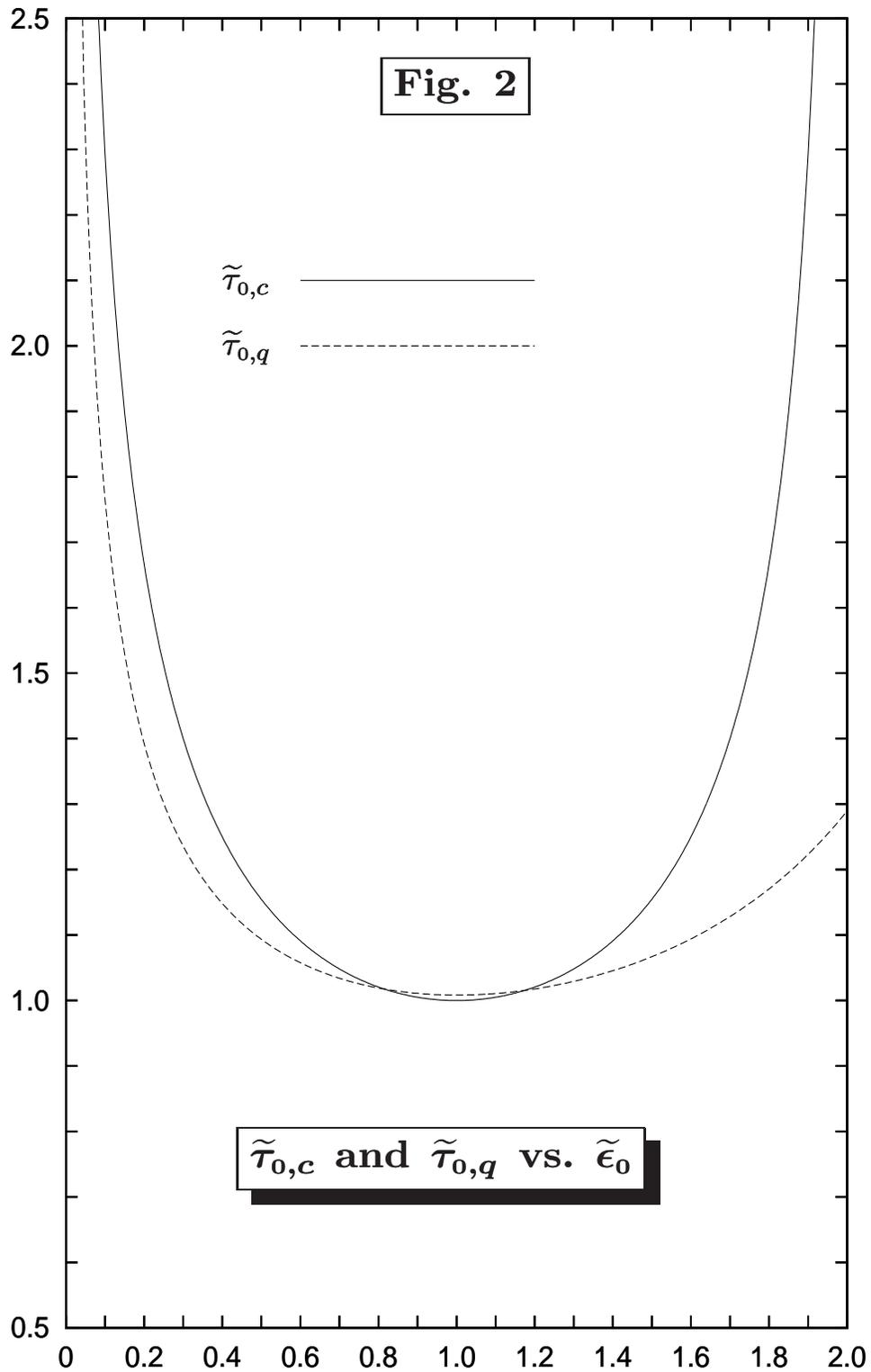}
\vspace*{-1.2cm}
 \caption{Delay times for complex and pure quaternionic  potentials in terms of the adimensional incoming energy $E_{\0}/\widetilde{E}_{\0}$. The difference between quaternonic and complex quantum mechanics is amplified for incoming energies closed to $2\,\widetilde{E}_{\0}$.}
\end{figure}

\end{document}